\documentclass[pre,aps,preprint,amsmath]{revtex4}

\usepackage{graphicx}% Include figure files
\usepackage{dcolumn}% Align table columns on decimal point
\usepackage{bm}% bold math

\begin{document}

\title{Intelligent thermal cloak-concentrators}

%\title{Intelligent Bifunctional Thermal Metamaterials}

%\title{Intelligent Bifunctional Thermal Metamaterials}

%\title{Combining Cloaks and Concentrators into One Device in a Mono-Physical Field;  Thermal Cloak-Concentrator with Intelligence}

%\title{Multifunctional Metamaterials under a Monophysical Field: Thermal Cloak-Concentrator}

%\title{Automatically switching from a concentrator to a cloak}

\author{Xiangying Shen$^{1}$}
\author{Ying Li$^{2}$}
\author{Jiping Huang$^{1,}$}\email{jphuang@fudan.edu.cn}
\author{Yushan Ni$^{2,}$}\email{niyushan@fudan.edu.cn}

\affiliation{$^1$Department of Physics, State Key Laboratory of Surface Physics, Key Laboratory of Micro and Nano Photonic Structures (Ministry of Education), and Collaborative Innovation Center of Advanced Microstructures, Fudan University, Shanghai 200433, China\\
$^2$Department of Mechanics and Engineering Science, Fudan University, Shanghai 200433, China}

\date{\today}% It is always \today, today,

\begin{abstract}
How to macroscopically control the flow of heat at will is up to now a challenge, which, however, is very important for human life since
heat flow is  a ubiquitous phenomenon in nature.  Inspired by intelligent electronic components or intelligent materials, here we demonstrate, analytically and numerically, a unique class of intelligent bifunctional thermal metamaterials called {\it thermal cloak-concentrators}, which can automatically change from a cloak (concentrator) to a concentrator (cloak) when the applied temperature field decreases (increases).  For future experimental realization, the behavior is also confirmed by assembling homogeneous isotropic materials according to the effective medium theory. The underlying mechanism originates from the effect of nonlinearity in thermal conduction. This work not only makes it possible to achieve  a switchable Seebeck effect, but also offers guidance both for macroscopic manipulation of  heat flow at will and for the design of similar intelligent multifunctional metamaterials in optics, electromagnetics, acoustics, or elastodynamics.
\end{abstract}

\maketitle
\section{Introduction} % (fold)
\label{sec:intro}

The flow of heat is a common phenomenon  in nature. However, how to macroscopically manipulate the heat flow  at will remains poorly explored in the literature, which, however, is very important for human life or military uses, say,   efficient refrigerators, solar cells,  energy-saving buildings, or infrared camouflage.
It is known that our world has been changing by {\it intelligent} electronic components or {\it intelligent} materials. Here the {\it intelligent} means that the component or material has properties that can be
tuned in a controllable manner by external stimuli. For example, electronic diodes belong to intelligent electronic components, which have much lower resistance to electric current along one  direction than the other; shape memory alloys are a kind of intelligent materials, which can recover their original shapes by  stimuli (say, heating~\cite{WenNC14} or adding an external magnetic field~\cite{PramanickPRL14}) after big deformations. Manipulating thermal conduction equations to achieve  desired temperature
distributions or heat flux patterns in thermal metamaterials is promising for  thermal extensions of {\it intelligent} electronic components or
{\it intelligent} materials. As a result, this manipulation might produce a unique class  of  intelligent thermal metamaterials, thus yielding a novel (intelligent) way for macroscopically controlling the flow of heat at will. For this purpose, let's trace back to the birth of transformation optics theory~\cite{leonhardtsci2006,pendrysci2006}.

%The recent years have witnessed the birth of cloaks or concentrators in these areas.

%Owing to nonlinearity, the device enables the amount of heat flux inside to quickly switch between zero (cloak) and a prominent value (concentrator).
In 2006, Leonhardt and Pendry {\it et al.} independently established the transformation optics theory, which leads to a new concept of invisibility cloaking~\cite{leonhardtsci2006,pendrysci2006}. Such a significant
progress soon enlightened a lot of scientists in different fields since it
offers a powerful method for designing metamaterials with new properties like optical illusions~\cite{Kadicrpp2013,Buckmannnat2014,Schittnysci2014}. In essence, this theory demands the form invariance of  domain equations
under linear coordinate transformations. This characteristic serves as the
foundation of transformation optics theory. Inspired by it, some researchers extended the transformation mapping theory to
the domain of thermal conduction, and then proposed a type of thermal
metamaterials, which are called thermal cloaks, for steady state heat flow
(i.e., the temperature, $T$, is independent of time) by using inhomogeneous anisotropic
 materials~\cite{fan2008shaped}. The original intention of designing such type of thermal metamaterials
is to hide an object inside the cloak from the detection through measuring the
external temperature distribution, and thus these metamaterials can thermally protect the central region containing the object.
So far, the steady-state thermal cloak~\cite{fan2008shaped} and its theoretical extensions~\cite{LiJAP10,Gueoe2012} have been
experimentally realized or developed~\cite{Narprl2012,SchittnyPRL13,XuPRL2014,HanPRL14,MaPRL14}.
Meanwhile, researchers also designed a lot of thermal metamaterials with
novel thermal characteristics beyond  cloaking~\cite{LiJAP10,Gueoe2012,HanEES13,Narprl2012,
Gaoepl2013,Gueoe2013,shen_thermally_2014,han_full_2014,mocciaprx2014,ZhuAIPA15,Chenepjap15,li_nonlinear_2015}, such as
concentrators~\cite{Narprl2012,Gueoe2012,HanEES13}. The concentrator helps to focus heat flux in a
particular region, and thus yields a higher temperature gradient inside the region, which can be used to improve the efficiency for thermal-to-electrical conversion (the Seebeck effect~\cite{WuPRL15,WangNC14}).

%It can be noted that the thermal cloak and concentrator are both useful in
%guiding heat flux.

%The possibility of manipulating thermal conduction to achieve the desired
%temperature distribution or heat flux pattern is promising
%for the development of thermal counterparts of intelligent electronic
%components.

Unfortunately, the existing thermal metamaterials published in the literature~\cite{fan2008shaped,Narprl2012,SchittnyPRL13,XuPRL2014,HanPRL14,MaPRL14,LiJAP10,Gueoe2012,HanEES13,
Gaoepl2013,Gueoe2013,shen_thermally_2014,han_full_2014,mocciaprx2014,ZhuAIPA15,Chenepjap15} don't have the
essential feature of  intelligence, namely, they lack the ability to
automatically sense and respond to the change of environmental temperature. The reason is
that within the framework of  transformation thermotics (or thermodynamics as also named by some other famous scholars)~\cite{fan2008shaped,Gueoe2012}, the property of thermal metamaterials can influence
the temperature field, but not vice versa due to  temperature-independent thermal conductivities under consideration. Evidently, it becomes necessary to develop the theory of nonlinear transformation
thermotics~\cite{li_nonlinear_2015}, which takes into account the fact that many materials have a temperature-dependent thermal conductivity. This theory suggests that new function can be
introduced into thermal metamaterials by using  a temperature-dependent transformation. In this work, we shall show that thermal metamaterials can be
made intelligent indeed based on the above idea by presenting a new thermal metamaterial --- intelligent thermal cloak-concentrator,
 which can automatically change from a cloak (concentrator) to a concentrator (cloak) when the applied temperature field decreases (increases).
%
%, and it can be employed to be building blocks in more complicated
%facilities in the future.
%

%It is a common idea that can come up with in one's mind --- a bi-functional
%material of concentration and cloaking. If the working temperature is low, the
%device can gather the heat in a certain region and when the temperature is
%getting higher, a thermally protection effect would be activated instead.

%However, it is obvious that the device demands nonlinear responses of
%consisting materials to the different temperatures, which is not discussed in
%original transformation mapping theory. As known to us, the thermal
%conductivities always change with temperature in the nature. Therefore, we
%have established the nonlinear transformation thermotics to serve the purpose.

%The method permits more flexibilities and possibilities in designing thermal
%metamaterials by tailoring the temperature dependent conductivities. We would
%like to show that by considering the nonlinear effects of materials, the
%mentioned switch between cloak and concentrator in one device can be achieved.

% section intro (end)

\section{Method} % (fold)
\label{sec:method}
The traditional thermal cloak helps to prevent heat from traveling inside the central region
without disturbing the temperature distribution outside, and thus thermally
hides an object inside it. As schematically drawn in Fig.~1(a), a
two-dimensional thermal cloak is designed according to a compression
transformation which maps a circle of radius $R_2$ to a ring with interior radius
$R_1$ and exterior radius $R_2$. In a polar coordinate system, the transformation
between original coordinates ($r,\theta$) and transformed ones ($r',\theta'$) is written as
\begin{equation}
	r'=r\dfrac{R_2-R_1}{R_2}+R_1, \theta'=\theta.
	\label{cloak1}
\end{equation}
According to this transformation, the thermal conductivity $\kappa$ of the
cloak is calculated from the equation of traditional transformation thermotics,
\begin{equation}
	\kappa=\dfrac{J\kappa_0 J^\mathrm{t}}{\det(J)},
	\label{kappa1}
\end{equation}
where $J$ is the Jacobian matrix of the transformation written in Cartesian
coordinate system, $\kappa_0$ is the thermal conductivity of the background,
and $\det(J)$ is the determinant of $J$.

The traditional thermal concentrator guides more heat flux into the central region and
increases the temperature gradient inside. As Fig.~1(b) shows, the
corresponding transformation of a two-dimensional concentrator maps a ring
with interior radius $R_3$ and exterior radius $R_2$ to a ring with interior radius
$R_1$ and exterior radius $R_2$. The formulas of this transformation in a polar
coordinate system are
\begin{equation}
	r'=r\dfrac{R_2-R_1}{R_2-R_3}+R_2\dfrac{R_1-R_3}{R_2-R_3}, \theta'=\theta.
	\label{concen1}
\end{equation}
The thermal conductivity of the concentrator can also be calculated by
substituting the Jacobian of Eq.~(\ref{concen1}) into Eq.~(\ref{kappa1}).

A comparison between Eq.~(\ref{cloak1}) and Eq.~(\ref{concen1}) reveals an
interesting fact: if we replace the $R_3$ in Eq.~(\ref{concen1}) with $0$, we
get exactly the transformation in Eq.~(\ref{cloak1}). The reason behind it is
that the original configuration of the transformation is a circle, and can
be regarded as a ring with interior radius $0$. This resemblance between the
principle of the cloak and the concentrator inspires us to combine these two
thermal metamaterials together and design a new kind of intelligent materials.
This device will automatically switches from a concentrator to a cloak or vice
versa as  environmental temperature changes. However, the traditional
transformation thermotics does not serve as a proper tool to design it because
the temperature won't affect the behavior of the device. Thus an extension of
the traditional theory is needed.

In our previous work~\cite{li_nonlinear_2015}, we've shown that the
transformation thermotics can be generalized by introducing the nonlinear
effect of temperature-dependent thermal conductivities. This theory enables us to make use of a certain temperature-dependent transformation
to achieve novel functions such as switchable thermal cloaking and macro thermal
diodes based on the cloaking~\cite{li_nonlinear_2015}. The foundation of this approach is the equivalence between such
transformation and a normal transformation applied on nonlinear materials (whose conductivity depends on  temperature).

Since we can change a concentrator into a cloak by simply replacing $R_3$ in
Eq.~(\ref{concen1}) with $0$, the cloak-concentrator can be designed on the basis  of a function $R^*(T)$ which switches between $R_3$ and $0$ at different temperature $T$.
Then the $T$-dependent transformation is
\begin{equation}
	r'=r\dfrac{R_2-R_1}{R_2-R^*(T)}+R_2\dfrac{R_1-R^*(T)}{R_2-R^*(T)}, \theta'=\theta.
	\label{concen-cloak}
\end{equation}
The thermal conductivity $\kappa(T)$ of our device is calculated
from a relation similar to Eq.~(\ref{kappa1}),
\begin{equation}
	\kappa(T)=\dfrac{J(T)\kappa_0 J^\mathrm{t}(T)}{\det(J(T))},
\end{equation}
where $J(T)$ is the Jacobian matrix corresponding to the transformation in
Eq.~(\ref{concen-cloak}) written in the Cartesian coordinate system. The
result expressed in the polar coordinate system is
\begin{equation}
	\kappa_r(T)=\kappa_0\left[{1+\dfrac{R_{2}(R^*(T)-R_1)}{r'(R_2-R^*(T))}}\right],
	 \kappa_{\theta}(T)=\kappa_0\left[{1+\dfrac{R_{2}(R^*(T)-R_1)}{r'(R_2-R^*(T))}}\right]^{-1},
	\label{kappa2}
\end{equation}
where $\kappa_r(T)$ and $\kappa_\theta(T)$ are the radial and azimuthal
components of $\kappa(T)$, respectively.

The specific term of function $R^*(T)$ is determined according to the desired
behavior of the cloak-concentrator. In this study, we want the device to behave
like a concentrator when the environmental temperature is lower than a critical
temperature, $T_c$, such that the central region inside it receives more heat flux from the
heat source. And when the environmental temperature increases to be higher than
$T_c$, we want the device to behave like a cloak in order to avoid heat flux
from coming inside the central region. In this case, an object located in the central region
is able to experience a relatively more stable thermal environment. To achieve
the above function, we resort to the following form of $R^*(T)$,
\begin{equation}
	%R^*(T)=R_3\left\{1-\left[1+\mathrm{e}^{\beta(T-T_c)}\right]^{-1}\right\},
    R^*(T)=R_3/\left[1+\mathrm{e}^{\beta(T-T_c)}\right],
	\label{R*}
\end{equation}
where $\beta$ is a scaling factor which determines the sensitivity of the
device to temperature.

% section method (end)
\section{Results} % (fold)
\label{sec:results}
In order to show the effect of the bifunctional device, we perform finite
element simulations based on the commercial software COMSOL Multiphysics to
see if the metamaterials work when the applied temperature fields change.
Without loss of generality, the units of length, temperature and thermal
conductivity are defined as $L_x/8$ ($L_x$ is the width of the simulation
box), the initial temperature and the conductivity of background, respectively.
Accordingly, the scales of the length, temperature and conductivity are
nondimensionalized throughout this work. An annulus with interior radius $R_1=1$ and exterior
radius $R_2=3$ is set in a box with size $8\times 7$ as shown in Fig.~2. The
thermal conductivities of the background and the central region are set as $\kappa_0=1$.
The thermal conductivity of the device is calculated according to Eq.~(\ref{kappa2}) for
 $R_3=2$. Heat
diffuses from the left boundary with high temperature $T_H$ to the right
boundary with low temperature $T_L$. Meanwhile, the upper and lower boundaries
of the simulation box are thermally isolated.

The simulation results of the switchable device are shown in Fig.~2(a,b) where the critical temperature $T_c$ is set to be $1.5$. In Fig.~2(a) for
low temperature ($T=1.2\sim1.4$), we observe that the concentrator is focusing the
heat flow into the central region and the temperature gradient is raised. On the
other hand, when the environmental temperature is high enough ($T=1.6\sim1.8$)
as in Fig.~3(b), the cloaking effect is ``turned on''. The heat flux in the
cloak are guided to detour around the core (namely, the central region) and the temperature inside this core is
turned to be a constant. Due to the antisymmetry, a switchable device between a cloak
in low temperature and a concentrator in high temperature can also be worked out
on the same footing.

Although the thermal conductivity of such a device is inhomogeneous and anisotropic
 [Fig.~2(a,b)], for future experimental realization we may resort to the effective medium theory (EMT)
\cite{Huangpr06,LiJAP10,Gaoprl10} which utilizes a laminated structure as an
approximation of anisotropic materials. According to the EMT, one layer contains two alternating homogeneous isotropic sub-layers of thicknesses $d_A$ and $d_B$ with
conductivities $\kappa_A(T)$ and $\kappa_B(T)$. In a polar coordinate system,
the effective parameters for a layer can be obtained from
\begin{equation}
%\begin{array}{lll}
\dfrac{1}{\kappa_{r}(T)} = \dfrac{1}{1+\eta}\left(\dfrac{1}{\kappa_A(T)}+\dfrac{\eta}{\kappa_B(T)}\right),
{\kappa}_{\theta}(T) = \dfrac{\kappa_A(T)+\eta\kappa_B(T)}{1+\eta},
\label{EMT1}
%\end{array}
\end{equation}
where $\eta=d_B/d_A$. In this work, we set $\eta = 1$ for simulations.

The method based on the EMT has been successful in the experimental demonstration of
thermal cloaks~\cite{Narprl2012,SchittnyPRL13}. For the case of  concentrators, the transformation is based on extensions; therefore, the array orientation of the
alternating sub-layers should be rotated with an angle of $\pi/2$. Considering both the aforementioned transformations, as shown in Fig.~2(c,d), we divide the device into
400 grids (which contain 5 layers consisting of 10 sub-layers in the radial direction
and 20 layers consisting of 40 sub-layers in the azimuthal direction).

Eq.~(\ref{EMT1}) is considered to be a reliable tool in realizing the
theoretical design, but the resulting forms of $\kappa_A(T)$ and $\kappa_B(T)$
obtained by solving  Eq.~(\ref{EMT1}) are too complicated. So in this study, we propose that instead of getting the
thermal conductivities directly from Eq.~(\ref{EMT1}), we solve two limiting
cases of concentrators and cloaks separately,
\begin{equation}
	\begin{split}
		\dfrac{1}{\kappa_{r}^{\mathrm{Co}}} &= \dfrac{1}{1+\eta}\left(\dfrac{1}{\kappa_A^{\mathrm{Co}}}+\dfrac{\eta}{\kappa_B^{\mathrm{Co}}}\right),
		{\kappa}_{\theta}^{\mathrm{Co}} = \dfrac{\kappa_A^{\mathrm{Co}}+\eta\kappa_B^{\mathrm{Co}}}{1+\eta};\\
		\dfrac{1}{\kappa_{r}^{\mathrm{Cl}}} &= \dfrac{1}{1+\eta}\left(\dfrac{1}{\kappa_A^{\mathrm{Cl}}}+\dfrac{\eta}{\kappa_B^{\mathrm{Cl}}}\right),
		{\kappa}_{\theta}^{\mathrm{Cl}} = \dfrac{\kappa_A^{\mathrm{Cl}}+\eta\kappa_B^{\mathrm{Cl}}}{1+\eta}.
	\end{split}
	\label{EMT2}
\end{equation}
In the above equations, the superscript, Co (or Cl), denotes the case of traditional
concentrators (or cloaks) without temperature dependency.
Then, according to the solutions of Eq.~(\ref{EMT2}), we find the desired relations of $\kappa_A(T)$ and $\kappa_B(T)$ with respect to temperature,
\begin{equation}
	\kappa_{A(B)}(T)=\kappa_{A(B)}^{\mathrm{Cl}} + \left(\kappa_{A(B)}^{\mathrm{Co}}-\kappa_{A(B)}^{\mathrm{Cl}}\right)\left[1+\mathrm{e}^{\beta(T-T_c)}\right]^{-1}.
	\label{EMT3}
\end{equation}

%After yielding $\kappa_A(T)$ and $\kappa_B(T)$,

Now we are
in a position to assemble the facility. Evidently, Fig.~2(c,d) displays the same
switching effect between concentrators and cloaks as presented in Fig.~2(a, b).

%are displayed .

Furthermore, for clarity, in Fig.~3, we choose the temperature gradient inside the device
to characterize the state of the intelligent thermal metamaterial. Since $\bm{q}=\kappa\bm{\nabla} T$, where $\bm{q}$ is the heat flux and $\kappa$ is the thermal conductivity of the central region ($\kappa=1$),
the temperature gradient represents the amount of heat flux in $x$-direction. A high value
of gradient in the central region indicates the device is working as a concentrator, while the near
zero value denotes that the cloaking effect is activated. For different temperature
fields applied, the variation of the temperature gradients are displayed
in Fig.~3. The results show that the new thermal metamaterial can make a
quick response to the change of environmental temperature, and the transition
process is sharp around the critical point $T_c$ ($T_c=1.5$). The simulation
results based on the EMT are also provided and the behaviors of the metamaterial
are in good agreement with the theory. The slight amount of temperature
gradient at the cloak side is due to a relatively few number of
layers (5 layers as plotted in Fig.~3) in the radial direction. This non-zero
temperature gradient can be further reduced to approach zero with more
alternative rings of thinner thickness. As an evidence, the concentrator
effects for the nonlinear transformation theory and EMT are very close to each
other because we used 20 layers of materials in the azimuthal direction (Fig.~3).
% section results (end)

\section{Discussion and Conclusion} % (fold)
\label{sec:conclusion}
By taking into account the role of nonlinearity in thermal conduction, we have designed a unique class of intelligent bifunctional thermal metamaterials ---
 thermal cloak-concentrators with intelligence. The cloak-concentrator is able to function as two traditional
metamaterials (namely, a cloak and a concentrator) at different environmental temperatures and quickly switches
from one to the other as temperature changes. With this new device, the heat
flux amount of a certain region can be automatically adjusted according to the
environmental temperature. For the convenience of future experimental demonstration, we have performed simulations based on the
 design of assembling homogeneous isotropic
materials  according to the EMT (effective medium theory); see Fig.~2(c,d). The
thermal conductivities  of theses materials vary with
temperature as determined by Eq.~(\ref{EMT3}), indicating that
the value of conductivities should increase or decrease rapidly with  temperature. Fortunately,
researchers have found several kinds of materials whose thermal conductivities
have this behavior~\cite{Zhengnat2011,Leenat2014}, which provide a viable
approach to realize the intelligent thermal cloak-concentrator proposed in this work.

Owing to the compression transformation of the original coordinate system,
the perfect cloaking effect provides a dead core inside the device, where the
temperature gradient are reduced to  zero. On the contrary, the
concentrator can raise the gradient in the central region due to the extension
transformation of the shell. Since for a given thermal conductivity the
temperature gradient determines the amount of heat flux, the thermal cloak can
significantly reduce the heat flux inside it, but the thermal concentrator
can raise the heat flux amount inside  to an abnormally high value.
Naturally, if one device can switch between a cloak and a concentrator, the
amount of its inner heat flux can vary in a wide range; if the switching
effect can be triggered by changing the environmental temperature, the
desired intelligence  is achieved.  Such a bifunctional thermal
metamaterial can automatically cause its central region to heat (or cool) in case of a low (or high)  environmental temperature.
 In particular, it can be used to switch on or off the Seebeck effect~\cite{WuPRL15,WangNC14} when the applied
temperature varies, thus achieving a so-called switchable Seebeck effect. Besides, it provides guidance for macroscopically  manipulating  heat flow at will in many thermal management problems, which are related to  heat preservation, dissipation, camouflage~\cite{han_full_2014} or illusions~\cite{ZhuAIPA15,Chenepjap15} in various kinds of areas ranging from energy-saving machines/buildings to military uses.
In addition, due to the similarity
between the governing equations, our consideration adopted in this work can be
extended to obtain the counterparts of such intelligent multifunctional thermal metamaterials in
other disciplines like optics, electromagnetics, acoustics, and elastodynamics.

%The cloak-concentrator suggests wide
%further applications in industry and civil life, such as energy-saving
%machines and heat preservation.

%This idea of combining different metamaterials
%together might serve as a convenient approach to design environment sensitive
%intelligent devices with multi-functions.

%It is up to now a challenge to design multifunctional metamaterials under a certain physical field in many disciplines like
%thermotics, electromagnetics, acoustics or elastodynamics.

{\it Acknowledgement.} The first two authors, X.S. and Y.L.,
contributed equally to this work. We acknowledge the financial support by the
National Natural Science Foundation of China under Grant Nos.~11222544 and
10576010 (Y.L. and Y.N.), by the Fok Ying Tung Education Foundation under Grant
No.~131008, by the Program for New Century Excellent Talents in University
(NCET-12-0121), and by the CNKBRSF under Grant No. 2011CB922004.

% section conclusion (end)

\clearpage
\newpage
Figure Captions

Fig.~1. Schematic graph showing  (a) a thermal cloak and (b) a thermal concentrator. Heat flow is represented by red lines with arrows.
The interior radius and exterior radius of the device is $R_1$ and $R_2$,
respectively. The area with radius $R_1$ denotes the central region or core.

Fig.~2. Finite-element simulations of the cloak-concentrator. (a,c) The device
functions as a concentrator at temperature below $1.4$ and (b,d) functions as
a cloak at temperature higher than $1.6$. The temperature distribution is
denoted by the color, with the white isothermal lines indicated. The left and
right boundaries are heat and cold sources with fixed temperatures. And the
upper and lower boundaries are thermally insulated. (a) and (b) show the results of Eq.~(\ref{kappa2}); (c) and (d)  display the results of Eq.~(\ref{EMT3}). Parameters: $R_1 = 1$, $R_2 = 3$, $R_3 = 2$, $\kappa_0=1$, and $T_c = 1.5$.

Fig.~3. Relation between the environmental temperature (defined as the average temperature of the heat and cold sources on the left and right boundaries) and the temperature gradient of the central region (which corresponds to the area of $r < R_1$ in Fig.~1) along $x$ direction. As shown in the figure, the red circles and blue squares denote the results based on the nonlinear transformation theory [Eq.~(\ref{kappa2})] and EMT [Eq.~(\ref{EMT3})] respectively.
The $x$-direction gradient outside the device keeps constant,
and it is indicated by the horizontal dashed line; the critical temperature $T_c$
is represented by the vertical dashed line.

%It should be noted that the transition process is sharp.

\clearpage
\newpage
\begin{figure*}[t]
\begin{center}
\centerline{\includegraphics[width=\linewidth]{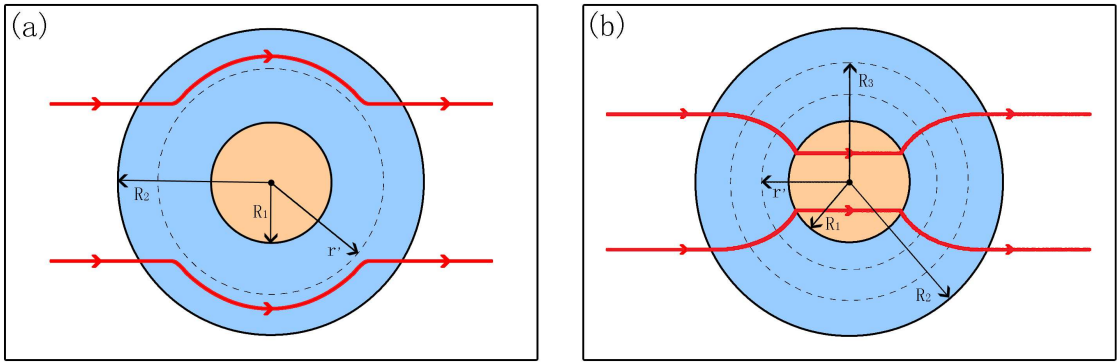}}
\caption{/Shen, Li, Huang, and Ni}
\end{center}
\end{figure*}

\begin{figure*}[t]
\begin{center}
\centerline{\includegraphics[width=\linewidth]{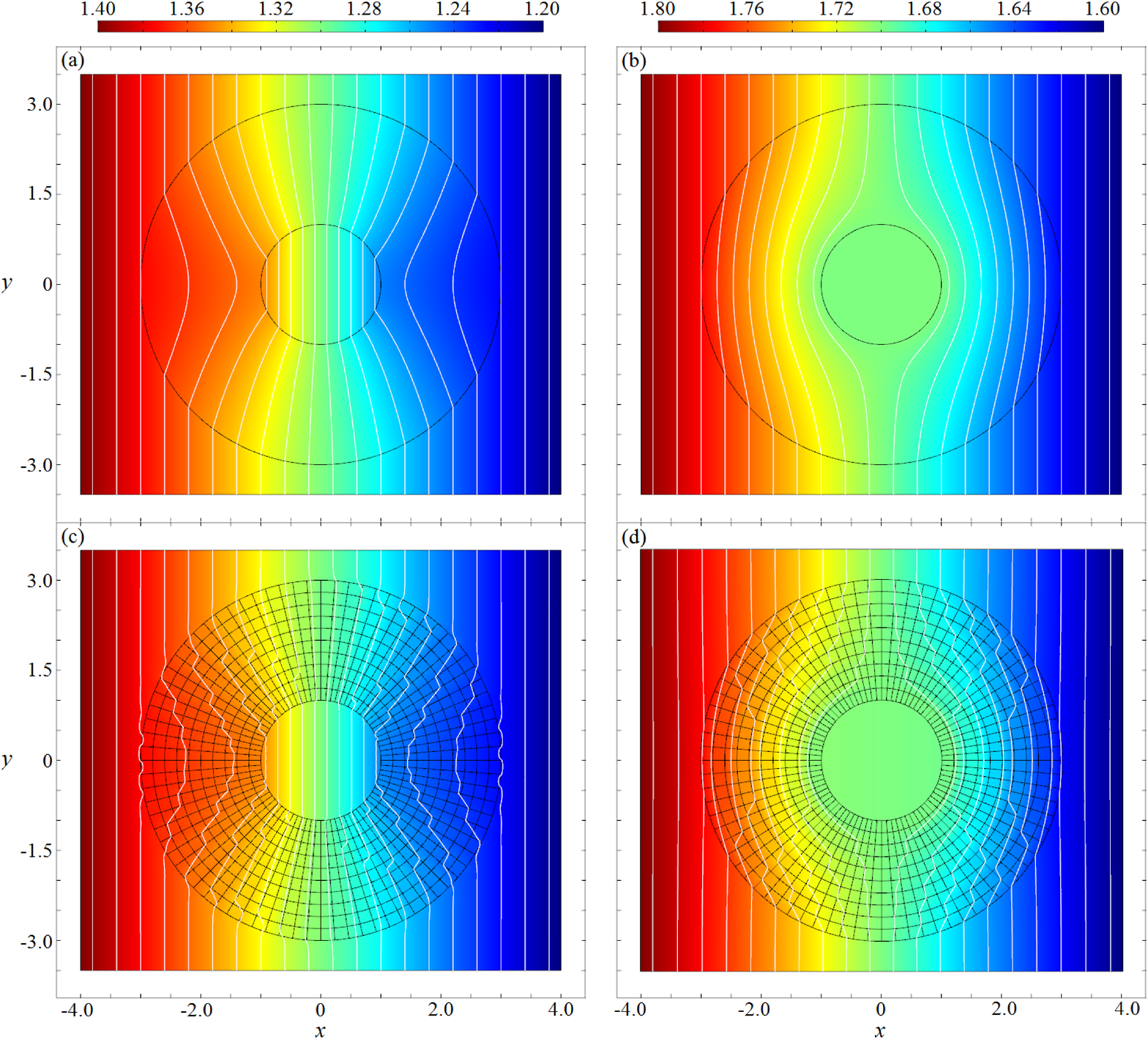}}
\caption{/Shen, Li, Huang, and Ni}
\end{center}
\end{figure*}

\clearpage
\newpage
\begin{figure*}[t]
\begin{center}
\centerline{\includegraphics[width=\linewidth]{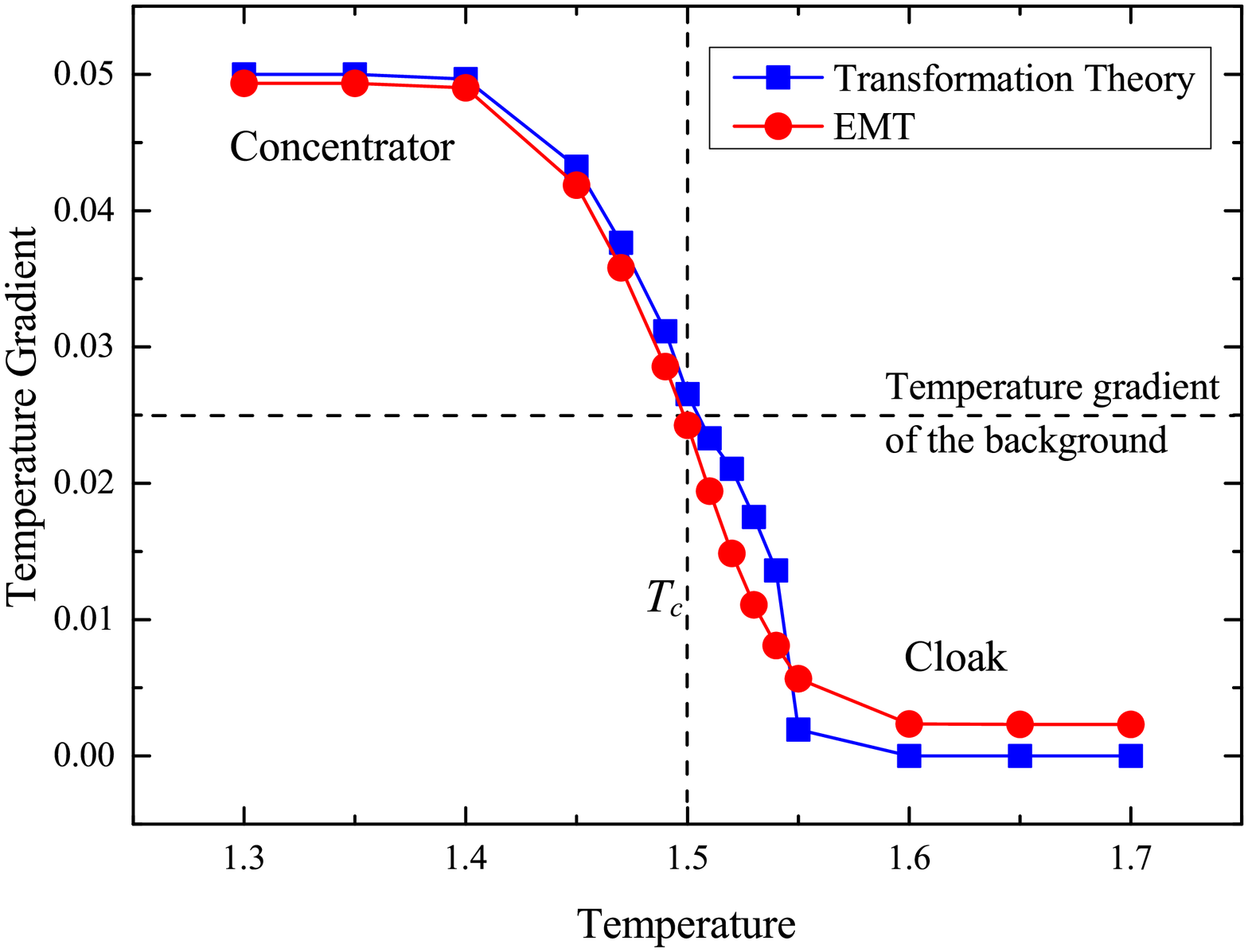}}
\caption{/Shen, Li, Huang, and Ni}
\end{center}
\end{figure*}
\end{document}